\newcommand{\uw}{U_{1\omega}}
\newcommand{\usw}{U_{2\omega}}
\newcommand{\sw}{S_{1\omega}}
\newcommand{\ssw}{S_{2\omega}}
\newcommand{\lw}{\ell_{1\omega}}
\newcommand{\lsw}{\ell_{2\omega}}
\newcommand{\kw}{k_{1\omega}}
\newcommand{\ksw}{k_{2\omega}}
\newcommand{\zw}{\zeta_{1\omega}}
\newcommand{\zsw}{\zeta_{2\omega}}
\newcommand{\nw}{n_{1\omega}}
\newcommand{\nsw}{n_{2\omega}}
\newcommand{\dnw}{\Delta n}
\newcommand{\dw}{D_{1\omega}}
\newcommand{\dsw}{D_{2\omega}}
\newcommand{\vw}{v_{1\omega}}
\newcommand{\vsw}{v_{2\omega}}
\newcommand{\rsw}{R_{2\omega}}
\newcommand{\tsw}{T_{2\omega}}

\documentclass[osajnl2,twocolumn,showpacs,superscriptaddress]{revtex4}  
\usepackage{array}
\begin{document}

\title{Experimental observation of second-harmonic generation and diffusion
inside random media}

\author{Sanli Faez}\email{faez@amolf.nl}
\affiliation{FOM Institute for Atomic and Molecular Physics AMOLF, Kruislaan
407, 1098 SJ Amsterdam, The Netherlands}
\author{P. M. Johnson}
\affiliation{FOM Institute for Atomic and Molecular Physics AMOLF, Kruislaan
407, 1098 SJ Amsterdam, The Netherlands}
\author{D. A. Mazurenko}
\affiliation{FOM Institute for Atomic and Molecular Physics AMOLF, Kruislaan
407, 1098 SJ Amsterdam, The Netherlands} \address{Zernike Institute for Advanced
Materials, University of Groningen, Nijenborgh 4, 9747 AG Groningen, The
Netherlands }
\author{Ad Lagendijk}
\affiliation{FOM Institute for Atomic and Molecular Physics AMOLF, Kruislaan
407, 1098 SJ Amsterdam, The Netherlands}

\begin{abstract}
We have experimentally measured the distribution of the second-harmonic
intensity that is generated inside a highly-scattering slab of porous gallium
phosphide. Two complementary techniques for determining the distribution are
used. First, the spatial distribution of second-harmonic light intensity at the
side of a cleaved slab has been recorded. Second, the total second-harmonic
radiation at each side of the slab has been measured for several samples at
various wavelengths. By combining these measurements with a diffusion model for
second-harmonic generation that incorporates extrapolated boundary conditions,
we present a consistent picture of the distribution of the second-harmonic
intensity inside the slab. We find that the ratio $\lsw/L_c$ of the mean free
path at the second-harmonic frequency to the coherence length, which was
suggested by some earlier calculations, cannot describe the second-harmonic
yield in our samples. For describing the total second-harmonic yield, our
experiments show that the scattering parameter at the fundamental frequency
$\kw\lw$ is the most relevant parameter in our type of samples.

\end{abstract}


\maketitle
\section{Introduction}
Optical second-harmonic generation inside inhomogeneous media has attracted much
interest for biological applications such as high-contrast
microscopy~\cite{campagnola-naturebio-03} and for photonic applications such as
high-efficiency frequency conversion in granular nonlinear
dielectrics~\cite{baudrier-nature-04}. The second-harmonic signal radiated from
a random medium is also an informative probe of the fundamental and the
second-harmonic intensity distribution inside the medium. Second-harmonic
signals provide valuable information about multiple-scattering processes inside
a random medium, which is not easily accessible by other methods. Previously,
these signals have been used to study the angular, spatial, and temporal
correlations inside scattering
media~\cite{deboer-prl-93,ito-pre-04,tomita-josab-05}.

Currently available first-principles theories of optical second-harmonic
generation in multiple scattering media consider only the case of the mean free
path being much larger than the wavelength, far from the Anderson localization
regime~\cite{kravtsov-prb-91,skipetrov-optcom-03}. Many articles have discussed
interference effects in multiple-scattering nonlinear media such as the effect
of weak localization or the enhanced forward scattering~\cite{kravtsov-prbr-91,
tiggelen-optcom-95, mahan-prb-98, wellens-pre-05}.

The efficiency of the nonlinear conversion in granular dielectrics is related to
the minimized role of phase-matching in such a medium. Phase-matching is
essential for the efficiency of three-wave mixing processes in nonlinear
materials. The optical dispersion in the nonlinear material results in a
phase-mismatch between the fundamental and the higher-harmonic propagating
light~\cite{boyd-book}.

There are two conventional methods to overcome the phase mismatch. The first
method makes use of the fact that in birefringent and nonlinear materials, it is
possible to fulfill the phase-matching condition by adjusting the crystal
orientation with respect to the beam. The second method, called
quasi-phase-matching, uses a periodically-polled polarizability of the material.
The nonlinearity of the crystal is modulated such that the extra momentum, which
results from the phase-mismatch, can be transferred to the crystal. In both of
these methods, the total conversion yield increases quadratically with the
length of light path inside the medium.

Recently, a novel method, called random-quasi-phase-matching, has been
introduced in which the crystal orientation is randomly varied along the beam
path by using a powder~\cite{baudrier-nature-04}. As suggested from theory and
confirmed by the experiment, for this new method, the accumulated
second-harmonic energy increases linearly with the sample thickness. By using
this method, a relatively high second-harmonic signal can be extracted from many
semiconductor powders, which were otherwise (in their bulk form) useless for
conventional methods. Using random-quasi-phase-matching, second-harmonic
generation can be obtained for a larger bandwidth and acceptance angle than
conventional methods.

A high second-harmonic yield has also been observed by Tiginyanu~\textit{et~al.}
for strongly-scattering porous-GaP~\cite{tiginyanu-apl-00}. Mel'nikov and
coworkers have shown that the second-harmonic signal in the specular-reflection
direction is enhanced by orders of magnitude after anodically-etching the
single-crystalline GaP wafer~\cite{melnikov-apb-04,golovan-apb-05}. However, no
comparison with a theoretical model had been provided for these observations.
The distribution of second-harmonic light inside such a medium has also not been
investigated experimentally.

In this article, we use two complementary techniques to determine the
distribution of the second-harmonic intensity inside a multiple-scattering slab.
In the first experiment, the spatial distribution of second-harmonic light
intensity at the side of a cleaved slab is measured. In the second experiment,
the total second-harmonic radiation at each side of the slab is measured. On the
macroscopic level, our experimental results confirm the predictions of the
diffusion theory for distribution of the second-harmonic intensity in a
multiple-scattering medium. On the microscopic level, however, the only
available theoretical model by Kravtsov~\textit{et~al.}~\cite{kravtsov-prb-91},
which is based on the diffusion theory, is unable to describe our experimental
results. Their model suggests the ratio $\lsw/L_c$ of the transport mean free
path at the second-harmonic frequency to the coherence length as the universal
parameter for the conversion yield. By performing measurements at a range of
scattering strengths and frequencies we find that their model does not apply to
our type of samples. Instead, our measurements show that the second-harmonic
yield follows a consistent dependence on the transport mean free path at the
fundamental frequency $\lw$.

In the following sections we briefly review the available theories of
second-harmonic generation in random media. We present our derivation of the
intensity distribution from a diffusion model with the extrapolated boundary
conditions. In the experimental section we describe two separate measurement
techniques which probe the distribution of second-harmonic light generation
inside a highly-scattering sample. Results of our experiments are then compared
with predictions of the diffusion model.

\section{Theory}
\subsection{Optical nonlinearity in opaque material}
An opaque medium may also be optically nonlinear. This nonlinearity can be an
intrinsic property of the bulk material or a result of the enormous interfacial
area present in porous objects. The second-order nonlinearity is absent in many
non-crystalline materials or crystal structures because of the presence of
inversion symmetry. Relatively large non-linearities may arise at the interfaces
of these materials with others or with vacuum due to symmetry-breaking at the
interface. The scientific understanding of optical nonlinear processes in
strongly-scattering materials is still in the preliminary phase. Some models
have been developed~\cite{kravtsov-prb-91, skipetrov-optcom-03} based on the
diffusion approximation, in which interference effects are assumed to be
averaged out and the sample size $L$ is much larger than the transport mean free
path.

In these diffusion models the incident wave at the fundamental frequency
$\omega$ experiences several scattering events before leaving the random medium.
Light at the second-harmonic frequency is generated during the
multiple-scattering process. The propagation direction of the second-harmonic
light is scrambled within one transport mean free path $\lsw$, thus becoming an
isotropic source of diffusive photons at the second-harmonic frequency. The
effect of phase-mismatch between the fundamental and the second-harmonic light
is negligible when the transport mean free path is much smaller than the
coherence length $L_c(\omega)\equiv{\pi}/{|2\kw-\ksw|}$, where $\kw$ and $\ksw$
are the wave-vector magnitudes at the fundamental and the second-harmonic
frequencies. Therefore, in a random medium that consists of grains showing
nonlinear response, the effect of constructive interference can be overcome by
selecting grain sizes to be smaller than the coherence length. It has been
experimentally shown that the second-harmonic yield from equal amount of
material increases with grain size until the grain size approaches the coherence
length~\cite{baudrier-nature-04}.

Overcoming the destructive interference due to phase-mismatch is also possible
by introducing scatterers inside a homogeneous nonlinear crystal. In such a
medium the fundamental and second-harmonic waves scatter differently, therefore
the destructive interference of the otherwise copropagating waves does not
occur, providing $\lw,\lsw \ll L_c$.

\subsection{\label{subsec:reviewtheory}Review of existing theories}
Second-harmonic generation in random media has been theoretically modeled for
three different systems in two reports. Here we briefly introduce and compare
these models as they will be needed for comparison with our experimental
results. The important prediction of these models is the magnitude of the
mesoscopic~\cite{endnote} second-harmonic conversion rate $\Gamma$ which is
defined as the second-harmonic intensity per volume, generated inside the
scattering medium, per unit of fundamental energy density squared .

In the first report Kravtsov~\textit{et~al.}~\cite{kravtsov-prb-91} have
considered two different systems. In the first system, scattering is introduced
in a slab of nonlinear dielectric by adding a small volume fraction of
point-like scatterers. The microscopic nonlinear polarizability is assumed to be
unaffected by the presence of scatterers, thus constant across the sample. We
refer to this model as ``homogeneous nonlinear background model''. The medium
can be optically dispersive, $\dnw\equiv\nsw-\nw\neq 0$. When $\dnw\leq 0$, the
conversion rate $\Gamma$ is independent of scattering parameters $\kw\lw$ and
$\ksw\lsw$. When $\dnw\geq 0$, $\Gamma$ is found to depend on $\kw\lw$ as
\begin{equation}\label{eq:shgcoefficient}
\Gamma= A\varepsilon_0\omega \nsw \nw^2 [\chi^{(2)}]^2
\;\textrm{arccot}[\frac{\pi\lsw}{(\frac{1}{2}+\frac{\lsw}{\lw})L_c}],
\end{equation}
where $A$ is a dimensionless prefactor, $\chi^{(2)}$ is the norm of the
nonlinear polarizability of the bulk medium averaged over all directions,  $L_c$
is the coherence length, and $\lw$ and $\lsw$ are mean free paths at fundamental
and second-harmonic frequencies. All parameters in Eq.~(\ref{eq:shgcoefficient})
can be frequency dependent due to the optical dispersion of the material. In
Eq.~(\ref{eq:shgcoefficient}), $\chi^{(2)}$ and $L_c$ are microscopic properties
of the bulk material while $\lw$ and $\lsw$ are mesoscopic properties that
depend on the porous structure of the material.
Equation~(\ref{eq:shgcoefficient}) predicts that as the mean free path at
second-harmonic frequency becomes small relative to the coherence length, the
conversion rate should increase. The derivation of this result assumes weak
multiple-scattering, $\kw\lw\gg1,\;\ksw\lsw\gg1$.

The second system that Kravtsov~\textit{et~al.}~\cite{kravtsov-prb-91} have
considered is a dense powder of nonlinear grains in which the grain size is
larger than the wavelength and much smaller than the mean free path. This system
is similar to samples analyzed in reference~\cite{baudrier-nature-04}. We call
this model ``nonlinear-powder model''.

In the second report, Makeev and Skipetrov~\cite{skipetrov-optcom-03} have
introduced a third system. They have modeled a suspension of colloidal
particles. We refer to this model as ``nonlinear colloidal suspension model''.
In their model, the conversion centers are the same as the scatterers and the
background medium is linear. They found that, for these suspensions, the
second-harmonic intensity divided by the number of scatterers shows no explicit
dependence on the multiple-scattering properties of the suspension. In their
model, the second-harmonic conversion rate is given by the following simple
product:
\begin{equation}\label{eq:shgcoefficient2}
\Gamma= B\rho\vsw\Sigma_{2\omega},
\end{equation}
where $\rho$ is the concentration of colloidal particles, $\Sigma_{2\omega}$ is
the total second-harmonic cross-section of an isolated particle, $\vsw$ is the
energy velocity at second-harmonic frequency, and $B$ is a dimensionless number.

In all three models, the nonlinearity is assumed to be so low that the
distribution of fundamental diffusive photons is not affected, therefore the
energy density at fundamental frequency $\uw(z,t)$ can be calculated by solving
the diffusion equation for an optically-linear multiple-scattering slab.

\subsection{\label{subsec:stationarytheory}Macroscopic distribution of the second-harmonic intensity inside an opaque slab}
Following the literature~\cite{kravtsov-prb-91, skipetrov-optcom-03}, we assume
the source distribution of the second-harmonic diffusive photons inside the
opaque material $\ssw({\mathbf {r}},t)$ to be equal to the conversion rate
$\Gamma$ times the square of the energy density of the fundamental light
$\uw({\mathbf {r}},t)$:
\begin{equation} \label{eq:shgsource}
\ssw({\mathbf {r}},t) = \Gamma \uw^2({\mathbf{r}},t).
\end{equation}
The exact value of $\Gamma$ depends on the model and the sample type
[Eq.~(\ref{eq:shgcoefficient}) sets an example]. Being a mesoscopic quantity,
$\Gamma$ will not affect the macroscopic intensity distribution except for
changing an overall prefactor.

For a slab of finite thickness, illuminated with a continuous plane wave of
intensity $I_0$, the following set of diffusion equations can be written for the
distribution of fundamental and second-harmonic diffusive-photon densities,
\begin{eqnarray}
\dw \frac{d^2 \uw(z)}{d z^2} =-\sw(z),\label{eq:slabfundamental}\\
\dsw \frac{d^2 \usw(z)}{d z^2} = -\Gamma \uw^2(z),\label{eq:slabshg}
\end{eqnarray}
where $\dw=\vw\lw/3$ and $\dsw=\vsw\lsw/3$ are the diffusion constants at
fundamental at second-harmonic frequencies, $\vw$ and $\vsw$ are the energy
velocities in the medium. The source term $\sw(z)$ denotes the distribution of
the diffusive light source at the fundamental frequency. These equations must be
solved together with the following boundary conditions~\cite{ishimaru-book},
\begin{eqnarray}
\uw(z)-\zw'\frac{d{\uw(z)}}{d{z}}=0 &\;\textrm{at}\,z=0, \label{eq:bcdiffusiona}\\
\uw(z)+\zw''\frac{d{\uw(z)}}{d{z}} = 0 &\;\textrm{at}\,z=L, \label{eq:bcdiffusionb}\\
\usw(z)-\zsw'\frac{d{\usw(z)}}{d{z}} = 0 &\;\textrm{at}\,z=0,\label{eq:bcdiffusionc}\\
\usw(z)+\zsw''\frac{d{\usw}(z)}{d{z}} =0
&\;\textrm{at}\,z=L,\label{eq:bcdiffusiond}
\end{eqnarray}
where $\zw',\,\zsw'$ are the extrapolation lengths at the incidence interface of
the slab, and $\zw'',\,\zsw''$ are the extrapolation lengths at the opposite
interface of the slab at the corresponding frequencies.

There are various ways of formulating the source distribution at fundamental
frequency $\sw(z)$. A phenomenological and perhaps the most practical
method~\cite{akkermans-prl-86}, is to set the source at one transport mean free
path from the incident interface, inside the slab: $\sw(z)=({\vw I_0}/{\lw})
\delta(z/\lw-1)$. A natural extension~\cite{durian-apop-95} to the previous
description considers a source term that exponentially decreases with depth,
with a decay length equal to the mean free path: $\sw(z)=({\vw
I_0}/{\lw})\exp(-z/\lw)$. In the following calculation we use the latter
description.

By integrating the diffusion equation~(\ref{eq:slabfundamental}) over an
exponential source and applying boundary conditions~(\ref{eq:bcdiffusiona})
and~(\ref{eq:bcdiffusionb}) we derive, for inside of the slab,
\begin{equation}\label{eq:distfundamental}
\uw(z)={3I_0}[\frac{(\lw+\zw')(L+\zw''-z)}{\lw(L+\zw'+\zw'')}
-\frac{2}{3}\exp(-\frac{z}{\lw})],
\end{equation}
where the thickness $L$ of the slab is considered much larger than the mean free
path, and thus terms such as $\exp(-L/\ell_{\omega})$ are neglected.

The second-harmonic energy density is given by inserting the result of
Eq.~(\ref{eq:distfundamental}) in diffusion equation~(\ref{eq:slabshg}) followed
by a double integration. The total generated second-harmonic radiation that is
propagating in the $2\pi$ spherical angle around the incidence direction is
referred to as the forward radiation and denoted by $\tsw$. The total radiation
that is propagating in the opposite (backward) direction, is denoted by $\rsw$.
For an slab illuminated by a plane wave, these two quantities are given by
\begin{eqnarray}\label{eq:rtdefinition}
\rsw&=&+\frac{\dsw}{\vsw}\frac{d\usw(z)}{dz}|_{z=0}\;, \\
\tsw&=&-\frac{\dsw}{\vsw}\frac{d\usw(z)}{dz}|_{z=L}\;.
\end{eqnarray}

Inserting the fundamental photon-density distribution~(\ref{eq:distfundamental})
in diffusion equation~(\ref{eq:slabshg}), and integrating yields the
second-harmonic photon-density distribution inside the slab,
\begin{equation}\label{eq:distshg}
\usw(z)= (\zsw'+z)\rsw-{\Gamma}\int_0^z \int_0^{z_1} \uw^2(z_2)dz_2dz_1.
\end{equation}
Boundary condition~(\ref{eq:bcdiffusionc}) has been applied. Applying the
boundary condition~(\ref{eq:bcdiffusiond}) yields $\rsw$. Expressions for
$\usw$, $\rsw$, and $\tsw$ as functions of $\Gamma, I_0, L, \lw, \zw,$ and
$\zsw$ were obtained by combining Eqs.~(\ref{eq:distfundamental}),
(\ref{eq:rtdefinition}), and~(\ref{eq:distshg}) with boundary
condition~(\ref{eq:bcdiffusiond}) in Mathematica. The closed forms of the
answers contain many terms and are not necessary to present here for the purpose
of the discussions in this paper. In completion of earlier
calculations~\cite{skipetrov-optcom-03}, we have included the extrapolated
boundary conditions and exponentially decaying source term for fundamental
light.

We have compared the internal distribution of second-harmonic intensity in the
limiting case of vanishing extrapolation length $\zw'=\zsw'=0$ with the
corresponding result stated in Ref.~\cite{skipetrov-optcom-03} and found that
their approach works well for the case of optically-thick slabs with zero
extrapolation length, i.e. absorbing boundaries. The effect of finite
extrapolation length is not negligible, regardless of the slab thickness.

Here we present the answer up to the first order in $\lw/L$. The first two
non-zero orders suffice for most of our discussions,
\begin{eqnarray}
\rsw&=&\frac{9\Gamma I_0^2
L}{4}(1+\frac{\zw'}{\lw})^2[1+\frac{4(2\zsw'+\zsw'')}{3L}-\\ \nonumber
&&\frac{8\lw^2(5\lw+6\zw')}{9L(\lw+\zw')^2}]+O(\frac{\lw}{L}), \label{eq:rtanswerr}\\
\tsw&=&\frac{3\Gamma I_0^2
L}{4}(1+\frac{\zw'}{\lw})^2(1+\frac{4(\zsw'+\zw'')}{L})+O(\frac{\lw}{L}).\nonumber\\
 \label{eq:rtanswert}
\end{eqnarray}

As has been mentioned earlier in this section, the mesoscopic conversion rate
$\Gamma$ shows up only in prefactors in
Eqs.~(\ref{eq:rtanswerr})~and~(\ref{eq:rtanswert}) and does not affect the
macroscopic distribution of radiation inside and around the slab. Our derivation
shows that the total second-harmonic intensity is enhanced with increasing
extrapolation ratio $\zw'/\lw$. Physically, this enhancement is caused by the
increased trapping of fundamental light inside the diffusing medium due to the
interfaces.

Due to the variability of several parameters from sample to sample, it turned
out to be useful to experimentally determine the ratio
\begin{equation}\label{eq:etadefined}
\eta\equiv\frac{\rsw}{\tsw}.
\end{equation}
In theory, the dimension-less quantity $\eta$ is independent of the
extrapolation ratio. Up to the the first order in $(\lw/L)$, $\eta$ is also
independent of the conversion rate $\Gamma$ and the incident intensity $I_0$. As
reported in earlier works~\cite{deboer-prl-93}, for optically-thick slabs, ratio
$\eta$ converges to a constant, i.e. $\lim\,\eta\rightarrow 3$ when
$L/\lw\rightarrow \infty$.

The most appropriate (currently available) theoretical model for our type of
samples is homogeneous nonlinear background model described in
Sec.~\ref{subsec:reviewtheory}, in which the conversion rate is given by
Eq.~(\ref{eq:shgcoefficient}). The intensity of the second-harmonic radiation is
dependent on several parameters. To compare any experimental result with any
multiple-scattering theory, it is advantageous to separate the dependencies
induced by multiple-scattering (mesoscopic) from variations that are caused by
intrinsic material properties (microscopic) and from geometrical specifications
of the sample (macroscopic). Therefore, for comparing the experimental results
with theory, we divide the measured second-harmonic backward radiation $\rsw$ by
the sample thickness, the incident intensity squared and all prefactors of the
arctan function in Eq.~(\ref{eq:shgcoefficient}), which are material properties
of GaP. We define the normalized yield as:
\begin{equation}\label{eq:normyield}
\gamma\equiv\frac{\rsw}{\omega \nsw \nw^2 [\chi^{(2)}]^2 L I_0^2},
\end{equation}
which is useful for comparing our data with the theories.

Note that in all of the above calculations, the absorption of fundamental and
second-harmonic light is neglected. This assumption is justified when the
absorption length of the medium is much larger than $L^2/\lw$, which a justified
assumption for our type of samples.

For a pulsed light-source, generally, the stationary calculation is valid if
$\tau_0\gg\tau_D$, where $\tau_0$ is the pulse duration and $\tau_D$ is the
Thouless time, defined as $\tau_D\equiv L^2/\dw$. Using analytical calculations,
which will be presented elsewhere, we have calculated the effect of pulse
duration on second-harmonic intensity distribution. Although the non-stationary
calculations are analytic, they contain long and complicated mathematical
expressions that do not provide any intuition about the physical outcome. We
found that for our specific experimental conditions, the difference between
non-stationary and stationary solutions is much less than our experimental
error, therefore we have presented our results only in comparison with the
stationary solutions.

\section{Samples and setup}\label{sec:setup}

\subsection{\label{subsec:sample}Samples}
The porous-GaP samples were etched from commercially available single crystal
$n$ type gallium phosphide wafers~\cite{erne-jes-96}. These wafers were doped
with donor atoms with a density range of $10^{18}\;\textrm{cm}^{-3}$. The
fabrication procedure is as follows. One side of a $9\times9\;\textrm{mm}^2$
piece of  GaP wafer with a thickness of 500 $\mu$m is anodically etched in
sulfuric acid under a controlled electromotive force. The electric current is
recorded in time. The total charge that is transported in the process is
proportional to the amount of removed material, allowing for the calculation of
the porosity for a known thickness of the etched layer. Different post-etching
processes can be used to remove the top layer, in order to improve optical
characterization, or to increase the pore-size. We used eight samples with
different values of mean free path. These samples were previously fabricated for
other optical experiments~\cite{schuurmans-science-99, rivas-thesis,
bret-thesis}.

\begin{table}
 \caption{\label{tbl:samplespec}Summary of specifications of the samples that
are analyzed in this report.$V_e$ is the applied electromotive force during the
etching process. The measured value of mean free paths $\lsw$ and $\lw$ are
presented at $\lambda_0=$ 0.65 and 1.3 $\mu$m, correspondingly. $L$ is the
thickness of the porous region.}
\begin{tabular}{c>{$}r<{$}>{$}c<{$}>{$}c<{$}>{$}c<{$}} \hline
\textrm{Tag}&L({\mu\mathrm{m}})&{\lsw({\mu\mathrm{m}})}&{\lw({\mu\mathrm{m}})}&V_e(V)\\
\hline
1&35.5\pm1&0.27\pm0.04&2.49\pm0.29&14.7\\
2&43.7\pm1&0.26\pm0.03&1.85\pm0.29&14.7\\
3&23.5\pm1&0.26\pm0.04&2.18\pm0.24&14.7\\
4&32\pm1&0.83\pm0.06&2.50\pm0.28&22.5\\
5&26\pm1&0.60\pm0.05&2.22\pm0.18&11.2\\
6&83\pm3&1.10\pm0.08&4.71\pm0.37&10.0\\
7&126\pm4&1.24\pm0.08&3.38\pm0.48&15.0\\
8&116\pm3&1.58\pm0.11&5.75\pm0.60&10.0\\
\hline
\end{tabular}
\end{table}

\begin{figure}
\includegraphics[width=3.5in]{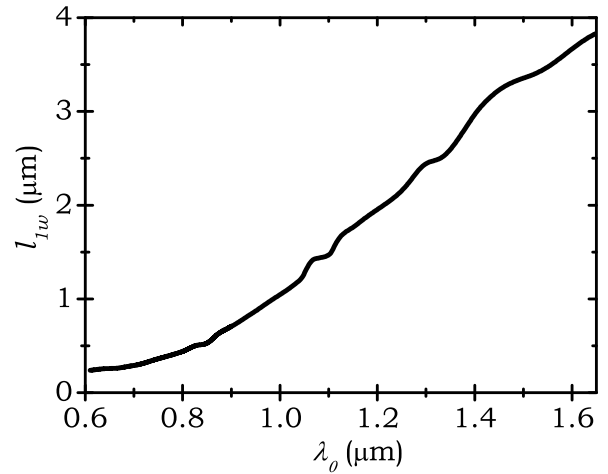}
\caption{\label{fig:mfp}The mean free path of a Porous-GaP sample is
wavelength-dependent. As a representative, we plot the mean free path as a
function of vacuum wavelength for sample 2. Similar curves are also obtained for
all samples using total-transmission measurements.}
\end{figure}

A summary of the scattering characterization properties of the investigated
samples is shown in table~1. The thicknesses were determined from scanning
electron microscope images. The mean free path values were determined by the
standard technique of measuring the linear total transmission. As a
representative, the mean free path of sample 2 has been determined over the
spectral range of 0.6 to 1.6 $\mu$m (Fig.~\ref{fig:mfp}) by coupling a
spectrometer to an integrating sphere setup. Dependence of the mean free path on
wavelength is largely due to the nano-structure of the samples. It also varies
from sample to sample. The magnitude of the nonlinear susceptibility of porous
GaP for some lattice orientations is comparable to that of a commercial
second-harmonic crystal such as potassium dihydrogen phosphate~(KDP). However,
as GaP is not birefringent, it has no application for conventional methods of
second-harmonic generation.

\subsection{\label{subsec:ratio}Setup for measuring total radiation}
\begin{figure}
\includegraphics[width=3.5in]{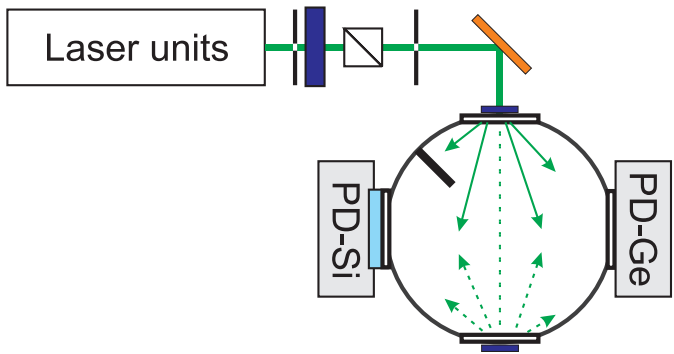}
\caption{\label{fig:setuptt}Experimental setup for measuring the total
second-harmonic radiation in forward and backward directions. The porous-GaP
slab is illuminated with a parallel beam of 150 femtosecond infrared pulses. The
integrating sphere collects the fundamental and the second-harmonic light
radiated in all directions from one side of the sample. The sample position is
either on the top of the sphere for transmission measurements or attached
beneath the sphere for reflection measurements. A Silicon photodetector (PD-Si)
behind a cold glass filter measures the second-harmonic signal in the visible
range. In parallel, a Germanium photodetector (PD-Ge) measures the transmitted
fundamental light at infrared range.}
\end{figure}
Fig.~\ref{fig:setuptt} shows a diagram of the integrating-sphere setup that was
used for measuring the total second-harmonic light that is radiated in the
forward or backward direction from a porous-GaP slab. Several gold mirrors guide
the beam vertically into the entrance of a TiO$_2$ coated integrating sphere.
The sample can be laid down at the entrance for transmission measurements or
placed beneath the sphere for reflection measurements. Two photodiodes are
positioned at two exit holes of the integrating sphere. At one exit-hole, an
amplified silicon detector (PDA55a - Thorlabs) connected to an oscilloscope
(DL9040L - Yokogawa) measures the second-harmonic signal after the fundamental
infrared light is filtered by a cold glass filter (KG5). At the other exit-hole,
a germanium detector (DET10A - Thorlabs) connected to another channel of the
same oscilloscope, measures the scattered fundamental light. In this way, both
fundamental and the second-harmonic signals can be measured simultaneously.

For a second-harmonic signal to be detectable by normal silicon detectors, an
incident intensity of several $\mathrm{GW/cm}^2$ is essential. A long pulse at
this intensity will damage GaP, therefore pulses shorter than nanoseconds with
low repetition rates are needed. The light source we used was a traveling-wave
collinear optical parametric amplifier of super-fluorescence (TOPAS - Light
Conversion) pumped with a sub-picosecond laser (Hurricane - Spectra Physics) at
800 nm and a repetition rate of 1 kHz. The output central wavelength could be
tuned continuously from 0.55 to 2.2 $\mu$m with a bandwidth of 5 nm. The pulse
was transform-limited with a duration of 150 fs. The undesired frequencies are
filtered out from the output beam by using a polarizer and a low-pass filter
(RG850 - Thorlabs).

For each sample, the radiated second harmonic light in the forward and backward
directions was measured in a fundamental-wavelength range of 1.2 to 1.6 $\mu$m.
Before each measurement, the incident power was checked by a pyroelectric head
and a power meter (Ophir). The transmittance of filters and responsivities and
the linearity of detectors were carefully taken into account before extracting
the second-harmonic yield.

Special attention is needed to correct for the transmittance of the non-porous
substrate. The total transmittance of the porous-bulk interface depends on the
refractive index mismatch at the interfaces and the directional distribution of
outgoing diffuse light known as the ``escape~function.'' We used the escape
function presented in~\cite{vera-pre-96}, with an effective refractive index of
$1.6 \pm 0.2$ for the porous region. This effective refractive index was
measured before based on the filling fractions and escape function measurements
from reference~\cite{rivas-thesis}. Using the Fresnel equations for
transmittance at substrate-air interface, the total angularly-averaged
transmittance of the GaP substrate for these samples has been calculated to be
$0.28\pm0.05$, for which the magnitude of the error is mainly due to the
uncertainty in the effective refractive index of the porous medium.

\subsection{\label{subsec:leakage}Setup for measuring effusion function}
\begin{figure}
\includegraphics[width=3.5in]{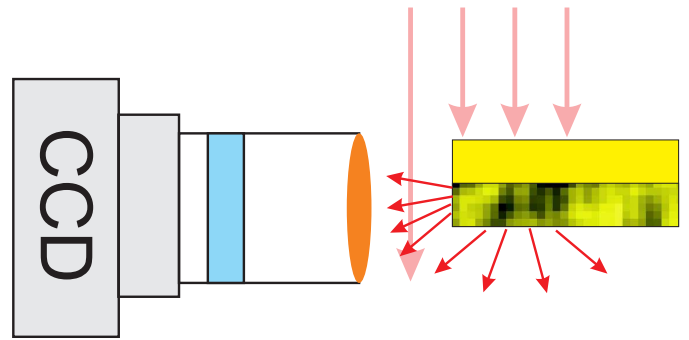}
\caption{\label{fig:setupef}Experimental setup for effusion microscopy
measurements. The sample consists of a thin porous and highly-scattering layer
laying on top of a transparent substrate. The CCD-camera images the
second-harmonic signal that is radiated from the narrow cross section of the
porous part of the sample while it is illuminated by a parallel beam of infrared
pulses.}
\end{figure}
In order to investigate the second-harmonic intensity distribution inside a
porous GaP sponge in greater detail, the second-harmonic radiation from the side
of a cleaved porous GaP slab was imaged under a microscope while the slab was
illuminated by a parallel beam of infrared pulses at normal incidence. We refer
to the diffusive energy flux at the interfaces of the random medium, which can
be measured with our method, as the ``\emph{effusion~function}". The measurement
of the effusion function is sensitive to details of the distribution of the
second-harmonic source inside the slab. In order to test this sensitivity, we
calculated the second-harmonic effusion function for a half-slab numerically and
compared it with the bulk second-harmonic energy distribution in a slab,
predicted by Eq.~(\ref{eq:distshg}). We observed that the second-harmonic
effusion pattern approximates the bulk second-harmonic intensity distribution
well. This result suggests that one can get an immediate qualitative sense of
the second-harmonic energy distribution from the observation of the
second-harmonic effusion function.

The effusion microscopy setup is shown in Fig.~\ref{fig:setupef}. The incident
beam in this setup is a Gaussian infrared beam of 3 mm in diameter with its
center aimed toward the edge of the sample. A cold glass filter in front of the
CCD camera blocks the fundamental light so that only the visible second-harmonic
light is captured.

\section{Results}

\subsection{\label{subsec:quadratic}Dependence on incident power}
\begin{figure}
\includegraphics[width=3.5in]{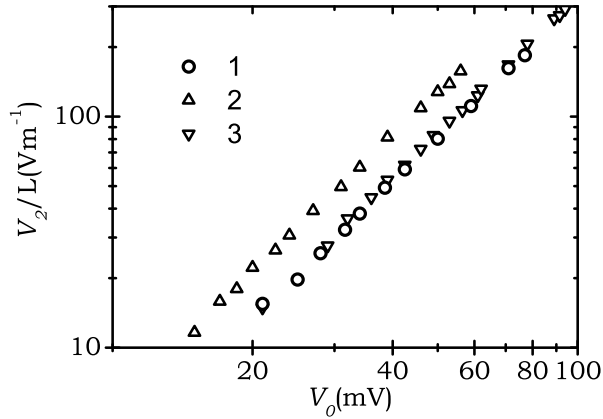}
\caption{\label{fig:quadratic}The backward-radiated second-harmonic signal
$V_{2}$, detected by the silicon detector, is scaled by the sample thickness and
plotted for three of the samples versus the signal detected by Germanium
detector $V_{0}$ in bi-logarithmic scales. The second-harmonic power is
proportional to $V_{2}$ and shows power-law dependence (exponent $=1.87\pm
0.03$) on the incident power, which is proportional to $V_{0}$. Numbers in the
legend indicate to sample tags as introduced in table 1.}
\end{figure}
At an incident wavelength of 1.2 $\mu$m, where the efficiency is the highest,
the dependence of second-harmonic yield on the incident power is measured for
several samples in the forward direction. The fundamental power is linearly
proportional to voltage $V_0$ of the silicon detector. The second-harmonic power
is linearly proportional to voltage $V_2$ of the germanium detector. The
measured relation between output voltages is plotted in
Fig.~\ref{fig:quadratic}. From this plot a consistent power-law dependence
between incident and second-harmonic powers is evident for these samples. The
result of fitting shows a power law $P_{2\omega}\propto P_0^{\alpha}$ with
$\alpha=1.87\pm0.03$. In a second-order nonlinear process the second-harmonic
intensity is proportional to the square of the incident intensity. The observed
deviation of the experimentally measured $\alpha$ from 2.0 may be a sign of
nonlinear (three photon) absorption inside the porous-GaP.

\subsection{\label{subsec:shgeffusion} Effusion function at second-harmonic frequency}
\begin{figure}
\includegraphics[width=3.5in]{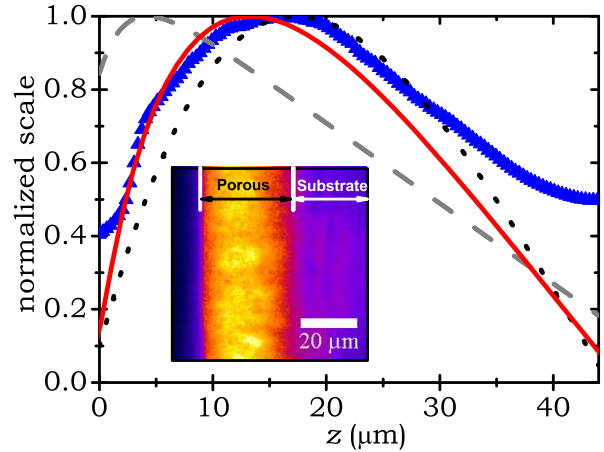}
\caption{\label{fig:shgleak}Inset: Micrograph of the second-harmonic effusion
intensity from a section of the porous-GaP slab while it is illuminated with a
parallel beam of infrared laser-pulses from left. Brighter regions indicate
higher effusion of second-harmonic light. Outset: The measured second-harmonic
intensity is averaged parallel to the substrate and its peak is normalized to
one. The result of the experiment (symbols) is plotted versus the position
inside the sample and is compared with the prediction of the stationary
diffusion model (solid line), found from the numeric calculation with no
adjustable fitting parameters. Our theoretical value for distribution of the
second-harmonic intensity in the bulk, which is presented in Sec. 2C, is shown
by the dotted line. The theoretical fundamental-frequency intensity distribution
inside the slab (dashed line) is plotted for comparison. }
\end{figure}

The microscopy setup described in Sec.~\ref{subsec:leakage} has been used to
capture the effusion function at the second-harmonic frequency. As the incident
beam size is much larger than the slab thickness, the effusion pattern is
laterally (parallel to the substrate) invariant. Roughness of the section due to
its porosity and cleaving-defects, can cause fluctuations in the observed
intensity. The intensity distribution is averaged in the lateral direction. The
experimental result is compared with our numeric calculation of the diffusion
model in Fig.~\ref{fig:shgleak}. The bulk distribution of the second-harmonic
intensity, is also plotted for comparison.

The measured second-harmonic intensity distribution qualitatively agrees with
the diffusion model. The intensity has a maximum around one third of the slab
thickness, as was predicted theoretically in~\cite{skipetrov-optcom-03}. The
largest deviation is observed near edges, which we attribute to the stray
second-harmonic light which is leaving the other interfaces of the slab behind
the imaging plane.

\subsection{\label{subsec:farfield}Second-harmonic intensity distribution in far-field}
\begin{figure}
\includegraphics[width=3.5in]{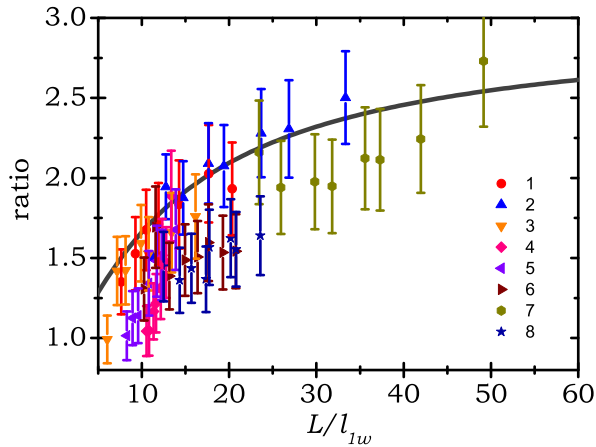}
\caption{\label{fig:universal}The ratio $\eta$ between total second-harmonic
light measured in the backward and the forward direction is plotted versus the
optical thickness $L/\ell _{1\omega }$ for various wavelengths and samples. The
stationary diffusion prediction from Sec. 2C is plotted as a solid line. We find
a good agreement between theory and measurements. Numbers 1-8 in the legend
correspond to the sample numbers introduced in table 1. }
\end{figure}

The measured value of $\eta$, defined by Eq.~(\ref{eq:etadefined}), has been
plotted versus the optical thickness of samples at various fundamental
wavelengths in Fig.~\ref{fig:universal}. The calculated $\eta$ in the framework
of stationary diffusion model, according to Eq.~(\ref{eq:etadefined}) is also
plotted for comparison. We find a good agreement between measurement and theory
for most of samples. Large error bars are mainly due to the uncertainty in the
effective refractive index of the porous medium.

\subsection{\label{subsec:efficiency}Second-harmonic yield}
At several incident frequencies, we have measured the total second-harmonic
intensity radiated from the sample. Extracting the absolute value of the
conversion rate from the second-harmonic intensity measurements is limited by a
large systematic shift which arises from uncertainties in the response function
of the setup. However, this systematic shift does not affect the relative
values. Therefore for testing theoretical models, we consider relative trends
rather than the absolute values.
\begin{figure}
\includegraphics[width=3.5in]{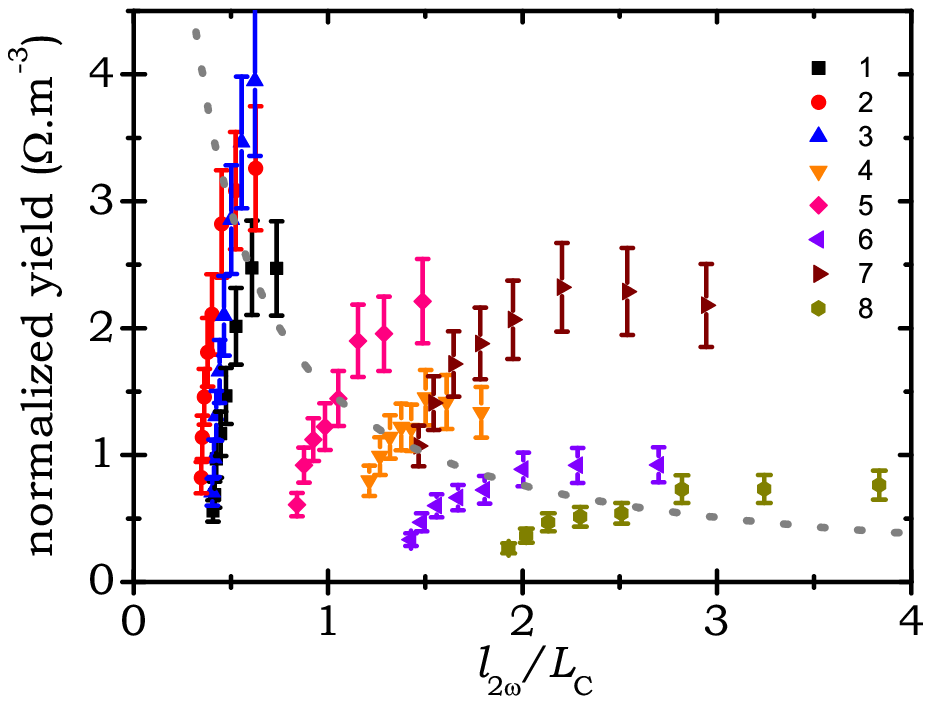}(a)
\includegraphics[width=3.5in]{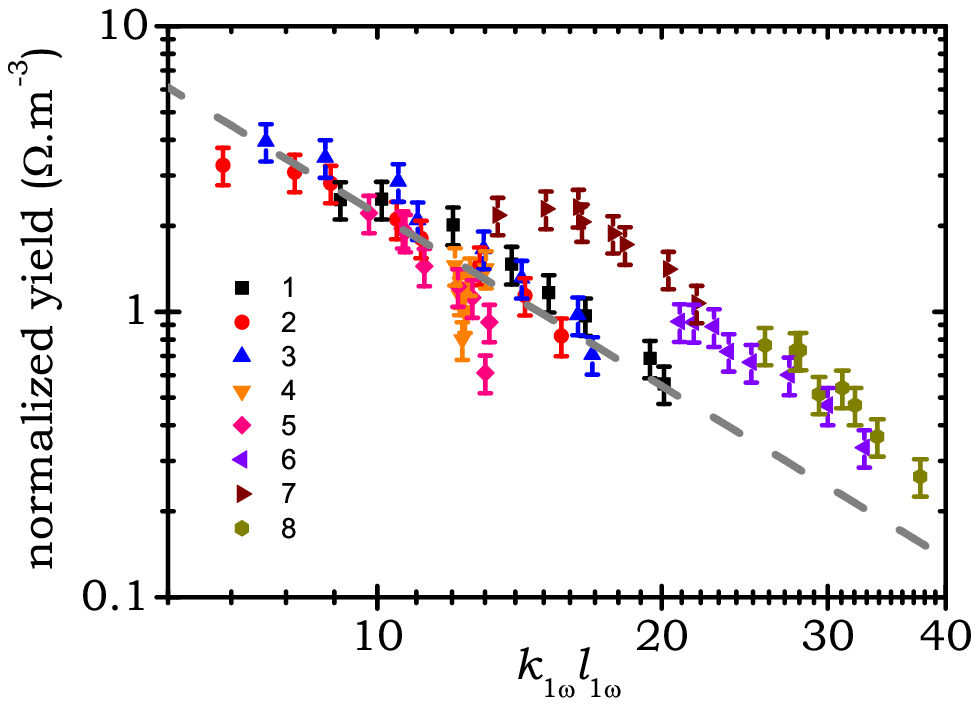}(b)
\caption{\label{fig:efficiency}(a)~The second-harmonic normalized yield as
defined in Eq.~(17) is plotted versus the ratio of the mean free path at second
harmonic frequency to the coherence length $\ell _{2\omega }/L_c$ for various
wavelengths and samples. The normalized yield is defined as the total
second-harmonic intensity generated in the backward direction divided by the
square of incident intensity and the thickness of the slab and normalized for
frequency dependent material properties of GaP. The dotted line shows the value
calculated from the theoretical model of
Kravtsov~{\em{et~al.}}~\cite{kravtsov-prb-91} plotted for comparison. No
agreement has been found between their theory and our measurements. Numbers 1-8
in the legend correspond to the sample numbers introduced in table~1. (b)~The
same data of (a) is plotted versus the scattering strength at the fundamental
frequency. The overall trend can be described by a power law relation, $\gamma
\propto (k_{1\omega }\ell _{1\omega })^\beta , \mskip \thickmuskip \beta
=-2.0\pm 0.3$ which is shown by the dashed line.}
\end{figure}

To check whether the homogeneous nonlinear background model of
Kravtsov~\emph{et~al.}~\cite{kravtsov-prb-91} can describe our data, the
normalized yield $\gamma$ defined by Eq.~(\ref{eq:normyield}) has been plotted
versus the ratio $\lsw/L_c$. This plot is shown in Fig.~\ref{fig:efficiency}(a)
for all samples and measured wavelengths. For this plot the optical dispersion
relation and the nonlinear polarizability of bulk GaP has been taken from
measurements of~\cite{aspnes-prb-83} and calculations of~\cite{levine-prb-94},
respectively. Our measurements show a trend opposite to the homogeneous
background model~\cite{kravtsov-prb-91}. For an individual sample, the yield
increases with increasing mean free path relative to the coherence length. The
incremental trend is different from sample to sample which indicates that
$\lsw/L_c$ is not the universal parameter for describing the conversion rate in
our kind of samples.

For the normalized yield plotted versus the scattering parameter at the
fundamental frequency $\kw\lw$, most of the measurement points for all the
samples are close to a single curve. The consistent trend of increasing yield
with decreasing $\kw\lw$ occurs both when comparing different wavelengths in a
single sample and when comparing different samples at the same wavelength. We
have fitted the data to a power law function and found an average exponent of
$-2.0\pm0.3$.

\section{Conclusion}
Using a microscopy technique, we have measured the second-harmonic intensity at
the side of a cleaved slab during its illumination with a Gaussian beam. We have
observed that the internal distribution of second-harmonic intensity predicted
by the diffusion model agrees with the experiment. The total intensity radiated
in the backward direction and the backward-forward ratio of the second-harmonic
intensity has also been measured for a number of samples. The measured
backward-forward ratios show good agreement with the results of the diffusion
model.

For describing our distribution measurements, we have presented a diffusion
model for second-harmonic intensity distribution in strongly-scattering
nonlinear material in which the diffused fundamental intensity is converted into
its second-harmonic via the process of degenerate two-photon mixing. The
second-harmonic light also diffuse in the scattering medium. The internal
distribution and the outgoing intensity in forward and backward directions of a
slab are derived based on the diffusion equation for light and extrapolated
boundary conditions. As has been reported previously~\cite{kravtsov-prb-91,
skipetrov-optcom-03}, diffusion theory predicts that in the slab geometry, the
total generated second-harmonic intensity increases linearly with the thickness
of the slab and depends quadratically on the extrapolation ratio at the
fundamental frequency. As a result of the diffusion model, the ratio $\eta$
between second-harmonic radiated intensity in backward and forward directions is
independent of the mesoscopic conversion efficiency and is only given by the
slab-geometry and boundary conditions. Therefore, $\eta$ provides a useful test
for the diffusion model, irrespective of the microscopic details. For a
sufficiently thick slab, $\eta$ approaches 3 as predicted by the diffusion
theory. This result has been confirmed by our measurements.

Although current diffusion models describe well the distribution of the
second-harmonic intensity, the overall conversion rate cannot be described by
the available theories. We have found a consistent power law dependence between
the second-harmonic yield and the scattering parameter at the fundamental
frequency, $\gamma\propto(\kw\lw)^\beta$, with an exponent of
$\beta=-2.0\pm0.3$.

\section{Acknowledgements}
We like to thank Willem Vos for sharing equipment, which was essential for our
experiments, and Otto Muskens for helpful collaboration and discussion. This
work is part of the research program of the ``Stichting voor Fundamenteel
Onderzoek der Materie'', which is financially supported by the ``Nederlandse
Organisatie voor Wetenschappelijk Onderzoek''.


\end{document}